\documentclass[11pt]{article}

\usepackage[brazil]{babel}
\usepackage{url}
\usepackage{amssymb,amsxtra} 
\usepackage{amsmath,color,graphicx}

\usepackage{graphicx}
\usepackage{indentfirst}
\RequirePackage{geometry}
\usepackage[T1]{fontenc}
\usepackage[bitstream-charter]{mathdesign}

\begin{document}

\def\chaptername{}
\def\contentsname{Sum\'{a}rio}
\def\listfigurename{Figuras}
\def\listtablename{Tabelas}
\def\abstractname{Resumo}
\def\appendixname{Ap\^{e}ndice}
\def\refname{\large Refer\^{e}ncias bibliogr\'{a}ficas}
\def\bibname{Bibliografia}
\def\indexname{\'{I}ndice remissivo}
\def\figurename{\small Fig.~}
\def\tablename{\small Tab.~}
\def\pagename{\small Pag.}
\def\seename{veja}
\def\alsoname{veja tamb\'em}


\setcounter{tocdepth}{3}

\clearpage
\pagenumbering{arabic}
\thispagestyle{empty}
\parskip 8pt

\vspace*{0.2cm}
\begin{center}
{\huge \bf O m\'{e}todo num\'{e}rico de Numerov}\\
\vspace*{0.3cm}
{\huge \bf aplicado \`{a} equa\c{c}\~{a}o de Schr\"{o}dinger}\\
\vspace*{0.2cm}
\textit{(Numerov numerical method applied to the Schr\"{o}dinger equation)}\\
\vspace*{1.5cm}
{\Large \bf \it Francisco Caruso$^{1,2}$ \& Vitor Oguri$^{2}$}\\*[2.em]

{{$^{1}$ Laborat\'{o}rio de F\'{\i}sica Experimental de Altas Energias\\ Centro Brasileiro de Pesquisas F\'{\i}sicas}}\\
{Rua Dr. Xavier Sigaud, 150 -- Urca, Rio de Janeiro, RJ -- 22290-180}\\*[2em]

{{$^{2}$ Instituto de F\'{\i}sica Armando Dias Tavares\\ Universidade do Estado do Rio de Janeiro}}\\
{Rua S\~{a}o Francisco Xavier, 524 -- Maracan\~{a}, Rio de Janeiro, RJ -- 20550-900}\\*[2em]

\end{center}






\vspace*{1.5cm}

\noindent \textbf{Resumo}

Neste artigo mostra-se como resolver numericamente problemas de autovalor associados a equa\c{c}\~{o}es diferenciais ordin\'{a}rias lineares de segunda ordem, contendo tamb\'{e}m termos que dependem da derivada primeira da vari\'{a}vel inc\'{o}gnita. Nesse sentido, faz-se uma apresenta\c{c}\~{a}o did\'{a}tica do m\'{e}todo de Numerov e, em seguida, ele \'{e} aplicado a dois pro\-blemas cl\'{a}ssicos da mec\^{a}nica qu\^{a}ntica n\~{a}o relativ\'{\i}stica cujas solu\c{c}\~{o}es anal\'{\i}ticas s\~{a}o bem conhecidas: o oscilador harm\^{o}nico simples e o \'{a}tomo de hidrog\^{e}nio. Os resultados num\'{e}ricos s\~{a}o confrontados com os obtidos analiticamente.\\
\textbf{Palavras-chave:} m\'{e}todo de Numerov, oscilador harm\^{o}nico, \'{a}tomo de hidrog\^{e}nio.

\vspace*{1.cm}

\noindent \textbf{Abstract}

In this paper it is shown how to solve numerically eigenvalue problems associated to second order linear ordinary differential equations, containing also terms which depend on the variable. A didactic presentation of the Numerov Method is given and, in the sequel, it is applied to two quantum non-relativistic problems with well known analytical solutions: the simple harmonic oscillator and the hydrogen atom. The numerical results are compared to those obtained analytically.\\
\textbf{Keywords:} Numerov method, harmonic oscillator, hydrogen atom.

\vfill

\newpage

\section{Introdu\c{c}\~{a}o} \label{introd}

A grande maioria dos m\'{e}todos num\'{e}ricos, como os de Newton, Euler, Lagrange, Gauss, Fourier, Jacobi, Runge-Kutta e tantos outros, foi introduzida no contexto das aplica\c{c}\~{o}es em f\'{\i}sica, astronomia ou em outras de natureza t\'{e}cnica, como na aerodin\^{a}mica \cite{Benzi}. Desde ent\~{a}o, a an\'{a}lise num\'{e}rica n\~{a}o era reconhecida como uma disciplina matem\'{a}tica e tal situa\c{c}\~{a}o perdurou durante as quatro primeiras d\'{e}cadas do s\'{e}culo XX. Hoje, apesar de alguns m\'{e}todos num\'{e}ricos serem ensinados nos cursos de f\'{\i}sica, no \^{a}mbito de  disciplinas da  matem\'{a}tica,   pouca \^{e}nfase \'{e} dada a eles nas aplica\c{c}\~{o}es f\'{\i}sicas. Com a populariza\c{c}\~{a}o dos computadores port\'{a}teis, cada vez mais acess\'{\i}veis ao grande p\'{u}blico, e capazes de  executar tarefas cada vez maiores e mais complexas, parece um contrassenso n\~{a}o explor\'{a}-los no ensino de f\'{\i}sica e de engenharia.

O matem\'{a}tico h\'{u}ngaro Peter Lax, do Instituto Courant da Universidade de Nova Iorque, refor\c{c}a a relev\^{a}ncia do ensino dos m\'{e}todos num\'{e}ricos destacando, com muita propriedade, seu aspecto universal e a import\^{a}ncia de os alunos explorarem solu\c{c}\~{o}es de equa\c{c}\~{o}es diferenciais utilizando computadores \cite{Lax}:

\begin{quotation}
\noindent M\'{e}todos num\'{e}ricos t\^{e}m a grande virtude que se aplicam universalmente. Quando s\~{a}o introduzidos m\'{e}todos especiais para lidar com a lamentavelmente pequena classe de equa\c{c}\~{o}es [diferenciais] que podem ser tratadas analiticamente, os alunos est\~{a}o aptos a perder de vista a ideia geral de que cada equa\c{c}\~{a}o diferencial tem uma solu\c{c}\~{a}o e que essa solu\c{c}\~{a}o \'{e} determinada unicamente pelos dados iniciais. Que hoje podemos utilizar computadores para explorar as solu\c{c}\~{o}es de equa\c{c}\~{o}es [diferenciais] \'{e} verdadeiramente revolucion\'{a}rio; estamos apenas come\c{c}ando a vislumbrar as consequ\^{e}ncias.\footnote{Numerical methods have the great virtue that they apply universally. When special methods are introduced to deal with each one of the pitifully small class of [differential] equations that can be handled analytically, students are apt to lose sight of the general idea that every differential equation has a solution and that this solution is uniquely determined by initial data. That today we can use computers to explore the solutions of [differential] equations is truly revolutionary; we are only beginning to glimpse the consequences.}
\end{quotation}

Na conflu\^{e}ncia dessas duas tend\^{e}ncias, procura-se divulgar aqui um poderoso m\'{e}todo de c\'{a}lculo num\'{e}rico desenvolvido originalmente por Boris Vasil'evich Numerov
\cite{Numerov_0,Numerov_1,Numerov_2}, aplicando-o \`{a}  equa\c{c}\~{a}o de Schr\"{o}dinger independente do tempo no caso de dois problemas t\'{\i}picos:  o oscilador harm\^{o}nico simples e o  \'{a}tomo de hidrog\^{e}nio. Esses s\~{a}o bons exemplos did\'{a}ticos, pois suas solu\c{c}\~{o}es anal\'{\i}ticas s\~{a}o bem conhecidas.

 O primeiro exemplo, para o qual se determina o espectro de energia do oscilador harm\^{o}nico simples como um problema de autovalor, e as respectivas autofun\c{c}\~{o}es $\psi(x)$, ou fun\c{c}\~{o}es de onda, ilustra o comportamento de uma part\'{\i}cula em um po\c{c}o de potencial unidimensional, quando a equa\c{c}\~{a}o de Schr\"{o}dinger n\~{a}o cont\'{e}m termos de derivada de primeira ordem.

No segundo exemplo, o qual envolve a equa\c{c}\~{a}o  de Schr\"{o}dinger contendo termo de derivada primeira, uma variante do m\'{e}todo original de Numerov ser\'{a} aplicada para se obter o espectro e a solu\c{c}\~{a}o radial do \'{a}tomo de hidrog\^{e}nio.

Com esses exemplos, ser\~{a}o esbo\c{c}ados alguns detalhes do procedimento geral utilizado no c\'{a}lculo da solu\c{c}\~{a}o num\'{e}rica de equa\c{c}\~{o}es diferenciais de segunda ordem.

\section{O m\'{e}todo de Numerov} \label{numerov}


A motiva\c{c}\~{a}o inicial de Numerov era poder calcular corre\c{c}\~{o}es \`{a} trajet\'{o}ria do cometa Halley. Portanto, na pr\'{a}tica, o \textit{m\'{e}todo de Numerov} foi desenvolvido, inicialmente, para determinar as solu\c{c}\~{o}es de problemas de autovalores associados a equa\c{c}\~{o}es diferenciais ordin\'{a}rias de
2$^{\rm a}$ ordem da mec\^{a}nica celeste, que n\~{a}o continham termos envolvendo a derivada
primeira de uma fun\c{c}\~{a}o inc\'{o}gnita $y(x)$, ou seja, equa\c{c}\~{o}es da forma
\begin{equation}\label{numerov_orig}
\frac{\mbox{d}^2y}{\mbox{d}x^2} = f(y,x) .
\end{equation}

Toda equa\c{c}\~{a}o do tipo~(\ref{numerov_orig}) pode ser substitu\'{\i}da pelo seguinte sistema de equa\c{c}\~{o}es de primeira ordem: \footnote{A Ref.~\cite{Allison} trata da solu\c{c}\~{a}o num\'{e}rica das equa\c{c}\~{o}es diferenciais acopladas de primeira ordem que resultam da equa\c{c}\~{a}o de Schr\"{o}dinger.}

$$\left\{
\begin{array}{l}
\displaystyle \frac{\mbox{d}z}{\mbox{d}x} = f(x,y)\\
\  \\
\displaystyle z = \frac{\mbox{d}y}{\mbox{d}x}
\end{array}
\right.
$$

Os m\'{e}todos tradicionais para resolver numericamente tal sistema de equa\c{c}\~{o}es, como os de Euler ou de Runge-Kutta, consideram que os valores de $y(x)$ e de $\mbox{d}y/\mbox{d}x$ sejam conhecidos em um dado ponto do dom\'{\i}nio $[a, b]$ de validade do sistema, \textit{i.e.}, s\~{a}o adequados para os chamados \textit{problemas de valor inicial}.

Em mec\^{a}nica qu\^{a}ntica n\~{a}o relativ\'{\i}stica, nos problemas de estados ligados que envolvem uma part\'{\i}cula de massa $m$ confinada  em um po\c{c}o de  potencial $V(x)$, em um dado intervalo $a <x <b$, as energias permitidas ($E$) e as correspondentes fun\c{c}\~{o}es de onda $\psi(x)$ que descrevem esses estados estacion\'{a}rios satisfazem a equa\c{c}\~{a}o de autovalor de Schr\"{o}dinger
\begin{equation}\label{Schr}
\frac{\mbox{d}^2\psi}{\mbox{d}x^2} + k^2(x) \psi = 0 ,
\end{equation}
em que $k = \sqrt{2m[E - V(x)]}/\hbar$\ e \ $\hbar \simeq 1,\!055 \times 10^{-34}$~J.s \'{e} a constante de Planck reduzida.

Nesses casos, como n\~{a}o se conhece o valor da derivada primeira da fun\c{c}\~{a}o de onda, os m\'{e}todos de Euler e Runge-Kutta n\~{a}o podem ser empregados. Entretanto, \'{e} poss\'{\i}vel estabelecer condi\c{c}\~{o}es de continuidade para os valores de $\psi$ e $\mbox{d}\psi/\mbox{d}x$ em dois ou mais pontos do dom\'{\i}nio da fun\c{c}\~{a}o de onda, o que caracte\-ri\-za os chamados \textit{problemas de valor de contorno}.

Al\'{e}m de tornar desnecess\'{a}ria a transforma\c{c}\~{a}o de uma equa\c{c}\~{a}o diferencial de segunda ordem em um sistema de primeira ordem, o m\'{e}todo de Numerov permite a determina\c{c}\~{a}o simult\^{a}nea do espectro de energia da part\'{\i}cula e das autofun\c{c}\~{o}es associadas a cada valor de energia.

Como todo m\'{e}todo num\'{e}rico iterativo, a solu\c{c}\~{a}o da equa\c{c}\~{a}o~(\ref{Schr}) \'{e} constru\'{\i}da por integra\c{c}\~{o}es sucessivas, realizadas passo a passo, a partir de valores
arbitr\'{a}rios para poss\'{\i}veis solu\c{c}\~{o}es em um ou mais pontos do dom\'{\i}nio de integra\c{c}\~{a}o.

Assim, no m\'{e}todo de Numerov, inicialmente, considera-se que a solu\c{c}\~{a}o seja conhecida em dois pontos subsequentes do intervalo $[a,b]$, por exemplo, em $\psi(x-\delta)$ e $\psi(x)$, sendo $\delta$ uma
quantidade arbitrariamente pequena, denominada \textit{passo} da integra\c{c}\~{a}o. A seguir, procura-se estabelecer, ent\~{a}o, um algoritmo num\'{e}rico para se  determinar a solu\c{c}\~{a}o no ponto seguinte, $\psi(x+\delta)$.

O ponto de partida para estabelecer esse algoritmo \'{e} a expans\~{a}o de $\psi(x\pm \delta)$ em  s\'{e}ries de Taylor, at\'{e} derivadas de quarta ordem, ou seja,
\begin{equation}\label{taylor}
 \psi(x \pm \delta) =  \psi(x) \pm \delta \psi^\prime (x) + \frac{\delta^2}{2}
\psi^{\prime\prime}(x)  \pm
\frac{\delta^3}{6} \psi^{\prime \prime \prime}(x)  + \frac{\delta^4}{24}\psi^{iv}(x) . \end{equation}

\noindent
Somando-se os termos $\psi(x + \delta)$ e $\psi(x - \delta)$, apenas as derivadas de ordem par sobrevivem e, portanto, chega-se a uma rela\c{c}\~{a}o entre os valores de uma fun\c{c}\~{a}o em tr\^{e}s pontos e sua derivada segunda, dada por
\begin{equation}\label{aprox-01}
\displaystyle \frac{\psi(x+\delta) + \psi(x-\delta) - 2 \psi(x)}{\delta^2} =
\psi^{\prime\prime}(x) + \frac{\delta^2}{12}\psi^{iv}(x)   \equiv
\left( 1 + \frac{\delta^2}{12} \frac{\mbox{d}^2}{\mbox{d}x^2} \right) \psi^{\prime\prime}(x) .
\end{equation}

Escrevendo a equa\c{c}\~{a}o de Schr\"{o}dinger unidimensional, equa\c{c}\~{a}o~(\ref{Schr}), na forma conveniente
\begin{equation}\label{aprox-02}
\left(1 + \frac{\delta^2}{12} \frac{\mbox{d}^2}{\mbox{d}x^2}
\right)\psi^{\prime\prime}(x) = - k^2(x) \psi(x) - \frac{\delta^2}{12}
\frac{\mbox{d}^2}{\mbox{d}x^2} \bigg[ k^2(x) \psi(x) \bigg] ,
\end{equation}
e utilizando a equa\c{c}\~{a}o~(\ref{aprox-01}) para substituir os termos que cont\^{e}m derivadas de segunda ordem, obt\'{e}m-se
\begin{eqnarray}\label{aprox-02a}
                  \nonumber \displaystyle && \frac{\psi(x+\delta) + \psi(x-\delta) - 2 \psi(x)}{\delta^2} \ =\ - k^2(x) \psi(x) - \frac{\delta^2}{12} \times\\
                  \nonumber \\
                   &\times& \left[\frac{k^2(x+\delta) \psi(x+\delta) + k^2(x-\delta) \psi(x-\delta) - 2 k^2(x) \psi(x)}{\delta^2} \right] + \mathcal{O} (\delta^4) .
\end{eqnarray}
Reagrupando a equa\c{c}\~{a}o~(\ref{aprox-02a}), obt\'{e}m-se a f\'{o}rmula de diferen\c{c}as de Numerov para o problema de uma part\'{\i}cula sob a\c{c}\~{a}o de um potencial unidimensional,

\begin{equation}\label{aprox-03}
\displaystyle
\left[1 + \frac{h^2}{12} k^2(x+\delta) \right]\psi(x+\delta) = 2 \left[1 -
\frac{5\delta^2}{12} k^2(x)\right] \psi(x) - \left[1 + \frac{\delta^2}{12}
k^2(x-\delta)\right] \psi(x-\delta) .
\end{equation}

Na realidade, cabe notar que o algoritmo pode ser aplicado a qualquer equa\c{c}\~{a}o diferencial ordin\'{a}ria linear e homog\^{e}nea de segunda ordem que n\~{a}o contenha termos de derivada primeira.

Uma vez que o problema de interesse \'{e} um problema de autovalor,  a t\'{e}cnica de integra\c{c}\~{a}o num\'{e}rica da equa\c{c}\~{a}o unidimensional de Schr\"{o}dinger para uma part\'{\i}cula em um po\c{c}o depende de se associarem
valores arbitr\'{a}rios convenientemente aos autovalores e \`{a}s respectivas (poss\'{\i}veis) autofun\c{c}\~{o}es em 2 pontos do dom\'{\i}nio do problema. Mas como
faz\^{e}-lo? Com rela\c{c}\~{a}o \`{a} escolha do valor inicial para a energia (primeiro autovalor), basta lembrar que, de acordo com a rela\c{c}\~{a}o de incerteza de Heisenberg, a
energia $E$ de uma part\'{\i}cula em um po\c{c}o de potencial $V(x)$ deve ser maior que o valor m\'{\i}nimo do po\c{c}o. Assim, considera-se, inicialmente, que $E_{\mbox{\tiny
inicial}} = V_{\mbox{\tiny min}} + \Delta E$, com $\Delta E > 0$.

\begin{figure}[hbtp]
\centerline{\includegraphics[height=6.0cm]{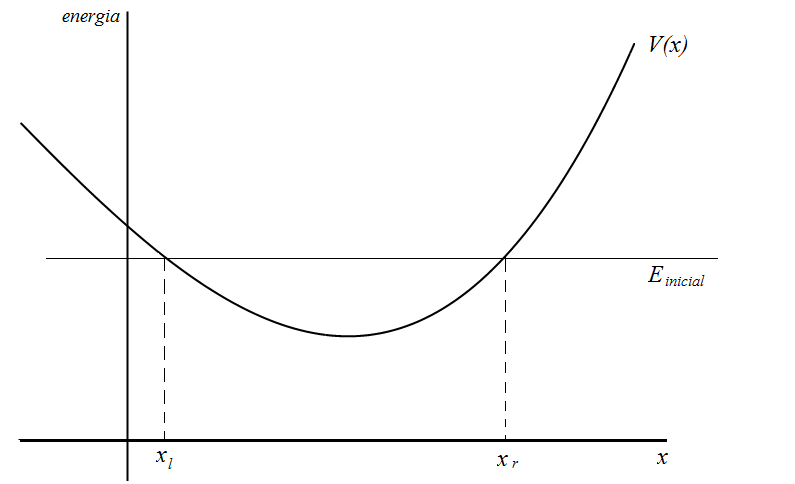}}
\vspace*{-0.35cm}
\caption{\small Curva de potencial.}
\label{osc}
\end{figure}

A escolha de um valor para a energia determina 2 pontos,  $x_{\ell}$ e $x_{r}$, nos quais o valor da energia \'{e} igual ao valor da energia potencial, e correspondem aos pontos de retrocesso de uma part\'{\i}cula cujo movimento obedece \`{a} mec\^{a}nica cl\'{a}ssica newtoniana.   Ou seja, do ponto de vista da mec\^{a}nica cl\'{a}ssica, o movimento da part\'{\i}cula est\'{a} restrito apenas \`{a} regi\~{a}o $[x_{\ell}, x_{r}]$, na qual a energia \'{e} maior ou igual \`{a} energia potencial.
As regi\~{o}es  $x <x_{\ell}$ \ e\  $x> x_{r}$ s\~{a}o denominadas regi\~{o}es classicamente proibidas.

Como a equa\c{c}\~{a}o de Schr\"{o}dinger admite solu\c{c}\~{o}es para essas regi\~{o}es classicamente proibidas, para cada valor de energia,
 inicialmente, se atribuiem valores para uma poss\'{\i}vel autofun\c{c}\~{a}o em 2 pontos das regi\~{o}es classicamente proibidas, nos quais  a fun\c{c}\~{a}o $\psi$ praticamente se anula. Em geral, esses s\~{a}o os pontos limites $a$ e $b>a$ do dom\'{\i}nio de integra\c{c}\~{a}o da fun\c{c}\~{a}o.

No entanto, a implementa\c{c}\~{a}o do m\'{e}todo de Numerov para a solu\c{c}\~{a}o do problema ainda requer um esquema de itera\c{c}\~{a}o que utiliza a f\'{o}rmula de Numerov em duas etapas: a  partir de $a$, ou \`{a} esquerda  de um dos pontos de retrocesso cl\'{a}ssico, doravante denominado {\it match point} ($x_{\mbox{\tiny  match}}$), e a partir de $b$, ou \`{a} direita  do {\it match point}.

Desse modo, tomando-se arbitrariamente um valor inicial para a energia, e dois valores arbitr\'{a}rios sucessivos para a solu\c{c}\~{a}o, a partir dos extremos inferior e superior do intervalo de integra\c{c}\~{a}o $[a,b]$, pode-se implementar o esquema de itera\c{c}\~{a}o do m\'{e}todo nos dois sentidos,  como:

\begin{enumerate}
\item Solu\c{c}\~{a}o \`{a} esquerda do {\it match point} ($ x < x_{\mbox{\tiny  match}}$). \\

Seja $E_{\mbox{\tiny inicial}} = V_{\mbox{\tiny min}} + \Delta E  \quad \Big(\Delta E/|V_{\mbox{\tiny min}}| \ll 1 \Big) $ \ um valor arbitr\'{a}rio  para a energia da part\'{\i}cula.
Arbitrando-se tamb\'{e}m valores para a fun\c{c}\~{a}o de onda, em 2 pontos sucessivos, a partir de $a$,
$$ \displaystyle
\left\{
\begin{array}{l}
\displaystyle
\psi^\ell (a) = 0 \\
\ \\
\psi^\ell (a + \delta) = \delta^\ell \qquad \qquad (\delta^\ell \ll 1)  \\
\end{array}
\right.
$$

e utilizando-se a f\'{o}rmula de diferen\c{c}as, equa\c{c}\~{a}o~(\ref{aprox-03}), a solu\c{c}\~{a}o \`{a} esquerda  \'{e} constru\'{\i}da
sequencialmente at\'{e} o {\it match point}  ($x_{\mbox{\tiny  match}}$), em que
$ \psi^\ell (x_{\mbox{\tiny  match}}) = \psi^\ell_{\mbox{\tiny  match}}$.

\item Solu\c{c}\~{a}o \`{a} direita do {\it match point} ($ x > x_{\mbox{\tiny  match}}$). \\

De maneira similar, para o mesmo $E_{\mbox{\tiny inicial}}$,  arbitrando-se
$$ \displaystyle
\left\{
\begin{array}{l}
\displaystyle
\psi^r (b) = 0 \\
\ \\
\psi^r (b - \delta) = \delta^r \qquad \qquad (\delta^r \ll 1)  \\
\end{array}
\right.
$$
\noindent a solu\c{c}\~{a}o \`{a} direita, a partir de $b$,  \'{e} constru\'{\i}da
sequencialmente at\'{e} os pontos $x_{\mbox{\tiny  match}}$ e $x =  x_{\mbox{\tiny  match}}  - \delta$, em que
 $$
 \left\{
 \begin{array}{l}
 \psi^r (x_{\mbox{\tiny  match}}) = \psi^r_{\mbox{\tiny  match}} \\
 \  \\
 \psi^r (x_{\mbox{\tiny  match}} - \delta) = \psi^r_{\mbox{\tiny  match} \scriptstyle - 1} \\
\end{array}
\right.
$$

\end{enumerate}

Para garantir a condi\c{c}\~{a}o de contorno da solu\c{c}\~{a}o, redefine-se a solu\c{c}\~{a}o \`{a} esquerda conforme a equa\c{c}\~{a}o~(\ref{continuidade}) dada a seguir, e testa-se a condi\c{c}\~{a}o de contorno das derivadas primeiras, segundo a equa\c{c}\~{a}o~(\ref{deriva_cont}).

 O procedimento \'{e} repetido passo a passo, nos dois sentidos, $a
\rightleftharpoons b$. Partindo-se de $a$,
  utilizando-se a f\'{o}rmula de recorr\^{e}ncia de Numerov associada \`{a} equa\c{c}\~{a}o, se
constr\'{o}i a solu\c{c}\~{a}o $\psi^\ell$ at\'{e} que se atinja o ponto de retrocesso
cl\'{a}ssico, por exemplo,  mais pr\'{o}ximo de $b$, no qual $E = V(x_{\rm match})$,
chamado de \textit{match point}. Depois, a partir de $b$,
faz-se o an\'{a}logo, construindo-se a solu\c{c}\~{a}o $\psi^r$ at\'{e} o \textit{match
point}. Em princ\'{\i}pio, as poss\'{\i}veis solu\c{c}\~{o}es $\psi^\ell$ e $\psi^r$ n\~{a}o
ser\~{a}o necessariamente iguais neste ponto $x_{\rm match}$. Para assegurar a continuidade
da solu\c{c}\~{a}o redefine-se $\psi^\ell$ como
\begin{equation}\label{continuidade}
\psi^\ell(x) \to \psi^\ell(x)  \, \frac{\psi^r (x_{\rm match})}{\psi^\ell (x_{\rm match})}  \qquad (a \leq x < x_{\mbox{\tiny  match}})
\end{equation}

Finalmente, verifica-se qu\~{a}o pr\'{o}ximos s\~{a}o os valores das respectivas derivadas primeiras de $\psi^r$ e da nova fun\c{c}\~{a}o $\psi^\ell$ assim
escalonada, no \textit{match point}.


Para se testar a condi\c{c}\~{a}o de contorno das derivadas primeiras,
tendo-se em conta as s\'{e}ries de Taylor para $\psi(x+\delta)$ e $\psi(x-\delta)$, at\'{e} a primeira ordem, pode-se escrever
\begin{equation}\label{deriva_cont}
\left\{
\begin{array}{l}
\displaystyle
 \frac{\mbox{d}\psi^\ell}{\mbox{d}x}\Big|_{x_r} =  \frac{\psi^\ell_{\rm{match}+1} - \psi^\ell_{\rm{match}-1}}{2\delta}  \\
 \ \\
\displaystyle
\frac{\mbox{d}\psi^r}{\mbox{d}x}\Big|_{x_r}  = \frac{\psi^r_{\rm{match}+1} - \psi^r_{\rm{match}-1}}{2\delta}
\end{array}
\right.
\end{equation}
nas quais $\psi_{\rm{match}\pm 1} = \psi (x_{\rm match}\pm \delta)$.

Se a diferen\c{c}a entre esses valores for menor que o valor de um erro predefinido, interrompe-se o processo, confirmando-se o autovalor
procurado e a respectiva autofun\c{c}\~{a}o como sendo

$$\psi(x) = \left\{\begin{array}{l}
 \psi^\ell (x) \qquad  \qquad (a \leq x < x_{\mbox{\tiny  match}})  \\
 \ \\
\psi^r (x)   \qquad \qquad ( x_{\mbox{\tiny  match}} \leq x \leq b)
\end{array}
\right.
$$

Se a condi\c{c}\~{a}o de continuidade das derivadas n\~{a}o for satisfeita,
incrementa-se o valor da energia para a busca de um novo valor, que seja
realmente um autovalor do problema, e de sua respectiva autofun\c{c}\~{a}o.

O processo pode ser repetido at\'{e} que se determine o n\'{u}mero desejado de
autovalores e autofun\c{c}\~{o}es do problema.

Por se basear na expans\~{a}o em s\'{e}rie de Taylor at\'{e} quarta ordem, o erro no m\'{e}todo de Numerov \cite{Blatt} \'{e} bem menor do que  o erro em m\'{e}todos  baseados em expans\~{a}o em ordem mais baixa, como o de Runge-Kutta.


\section{O oscilador harm\^{o}nico}

A equa\c{c}\~{a}o de Schr\"{o}dinger para uma part\'{\i}cula de massa $m$ em um campo
conservativo, tal como um po\c{c}o de potencial unidimensional $V(x)$,
pode ser escrita como uma equa\c{c}\~{a}o de autovalor
$$ H \psi (x) = E \psi (x) , $$
na qual as energias $E$ s\~{a}o os autovalores e as fun\c{c}\~{o}es $\psi (x)$, as respectivas
autofun\c{c}\~{o}es de quadrado integr\'{a}vel do
operador hamiltoniano $H$ dado por
$$H = -\frac{\hbar^2}{2m}\, \frac{\mbox{d}^2}{\mbox{d}x^2} + V(x) . $$

Considerando que um potencial f\'{\i}sico real se anula no infinito, as
autofun\c{c}\~{o}es tamb\'{e}m se  anulam nos
 extremos de um dado intervalo $[a,b]$ que ser\'{a} o dom\'{\i}nio de integra\c{c}\~{a}o da equa\c{c}\~{a}o,
  \textit{i.e.}, $\psi(a) = \psi(b)
= 0$.

Assim, de um outro ponto de vista, o problema do po\c{c}o unidimensional \'{e} um problema de autovalor
que envolve uma equa\c{c}\~{a}o diferencial linear e homog\^{e}nea de segunda ordem do tipo
\begin{equation} \label{Schr-eq}
\psi^{\prime \prime}(x) =  - k^2(x)\,  \psi(x) ,
\end{equation}
sujeita a condi\c{c}\~{o}es de contorno nos limites de um intervalo $[a,b]$, sendo
$$\displaystyle k^2(x) = \frac{2m}{\hbar^2} \Big[ E - V(x) \Big] . $$

Um exemplo t\'{\i}pico de um problema em mec\^{a}nica qu\^{a}ntica, que envolve a equa\c{c}\~{a}o de Schr\"{o}dinger sem termo de derivada primeira,  \'{e} a determina\c{c}\~{a}o dos autovalores e autofun\c{c}\~{o}es do oscilador harm\^{o}nico de massa $m$ e frequ\^{e}ncia natural $\omega$, cuja energia potencial \'{e} dada por
$$V(x) = \frac{1}{2} m\omega^2 x^2 . $$

Assim, $k^2(x) $ pode ser escrito como
$$\displaystyle k^2(x) = \frac{2m}{\hbar^2} \left[ E - \frac{1}{2} m \omega^2 x^2  \right] = 2 \frac{m\omega}{\hbar}
\left[ \frac{E}{\hbar \omega} -\frac{1}{2} \left( \frac{m \omega}{\hbar}\right) x^2 \right] , $$
e a equa\c{c}\~{a}o de Schr\"{o}dinger pode ser escrita como
$$\psi^{\prime \prime} (x) = 2 \left( \epsilon -\frac{1}{2}  x^2 \right) \psi(x) , $$
onde $x$ est\'{a} em unidades de $\sqrt{m\omega/\hbar}$ e a energia $\epsilon$, em unidades de $\hbar \omega$.

\vspace{0.2cm}
A Tabela~\ref{osc-tab} mostra a compara\c{c}\~{a}o dos 6 primeiros  autovalores ($\epsilon$) de um  oscilador harm\^{o}nico,  calculados anal\'{\i}tica e numericamente, em unidades de $\hbar \omega$,  e a figura~\ref{osc}, as correspondentes autofun\c{c}\~{o}es,  numericamente determinadas,  em fun\c{c}\~{a}o de $x$, onde $x$ est\'{a} em unidades de $\sqrt{m\omega/\hbar}$.

\renewcommand{\arraystretch}{1.3}
\vspace{-0.3cm}
\begin{table}[hbtp]
  \caption{\small Compara\c{c}\~{a}o do espectro de energia do oscilador harm\^{o}nico anal\'{\i}tica e numericamente calculado.}
  \centering
  \begin{tabular}{c|c|c}
  \hline
~~~$\epsilon(E/\hbar \omega)$ ~~~ & ~~~ anal\'{\i}tico~~~ & ~~~ num\'{e}rico~~~\\
\hline
  $\epsilon_0$  & 0,5 & 0,5\\
  $\epsilon_1$  & 1,5 & 1,5\\
$\epsilon_2$ & 2,5 & 2,5\\
$\epsilon_3$  & 3,5 &  3,5 \\
$\epsilon_4$  & 4,5 &  4,5 \\
$\epsilon_5$  & 5,5  & 5,5 \\
  \hline
  \end{tabular}
  \label{osc-tab}
\end{table}
\renewcommand{\arraystretch}{1.0}

\begin{figure}[hbtp]
\vspace{-0.2cm}
\centerline{\includegraphics[height=7.0cm]{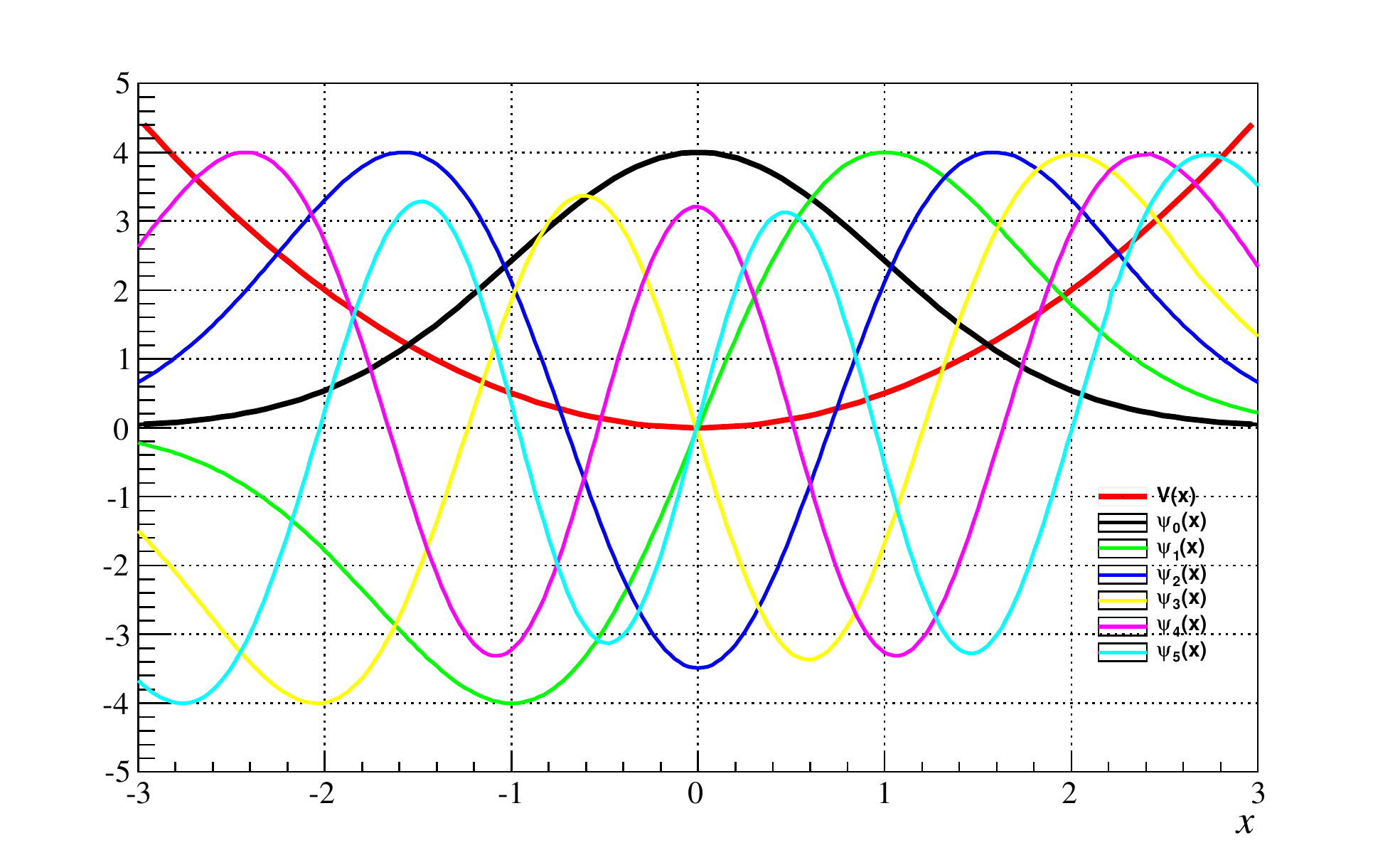}}
\vspace*{-0.35cm}
\caption{\small Energia potencial (em vermelho) e as autofun\c{c}\~{o}es do oscilador harm\^{o}nico para os seis primeiros n\'{\i}veis de energia.}
\label{osc}
\end{figure}


\section{O \'{a}tomo de hidrog\^{e}nio n\~{a}o-relativ\'{\i}stico}

Apesar de ter sido desenvolvido para equa\c{c}\~{o}es diferenciais ordin\'{a}rias lineares e homog\^{e}neas de segunda ordem que n\~{a}o contenham termos de derivada primeira, o m\'{e}todo de Numerov pode ser generalizado para abranger a presen\c{c}a de termos que contenham derivada primeira na equa\c{c}\~{a}o diferencial,  de modo a se poder considerar tamb\'{e}m problemas de autovalor \cite{Leroy}.

De fato, no caso de equa\c{c}\~{o}es lineares, toda equa\c{c}\~{a}o diferencial de segunda ordem do tipo
$$ \frac{\mbox{d}^2y}{\mbox{d}x^2} + P(x) \frac{\mbox{d}y}{\mbox{d}x} + Q(x) y = 0 ,$$
pode ser escrita em sua \textit{forma normal} \cite{Simmons}
$$ \frac{\mbox{d}^2y}{\mbox{d}x^2} + q(x) y = 0 ,$$
onde
$$ q(x) = Q(x) - \frac{1}{4} P^2(x) - \frac{1}{2} \frac{\mbox{d}P}{\mbox{d}x} .$$

A equa\c{c}\~{a}o radial de Schr\"{o}dinger para uma part\'{\i}cula de massa $m$ sob a a\c{c}\~{a}o de um campo  el\'{e}trico coulombiano, como o el\'{e}tron no \'{a}tomo de hidrog\^{e}nio,  pode ser escrita como

\begin{equation}\label{eq-radial-3d}
\frac{\mbox{d}^2 R(r)}{\mbox{d} r^2} + \frac{2}{r}
\frac{\mbox{d}R(r)}{\mbox{d}r} + \frac{2m}{\hbar^2} \left[ E + \frac{e^2}{r}  -
\frac{\hbar^2}{2m} \frac{\ell(\ell+1)}{r^2} \right] R(r) = 0 .
\end{equation}

\begin{figure}[hbtp]
\centerline{\includegraphics[height=7.0cm]{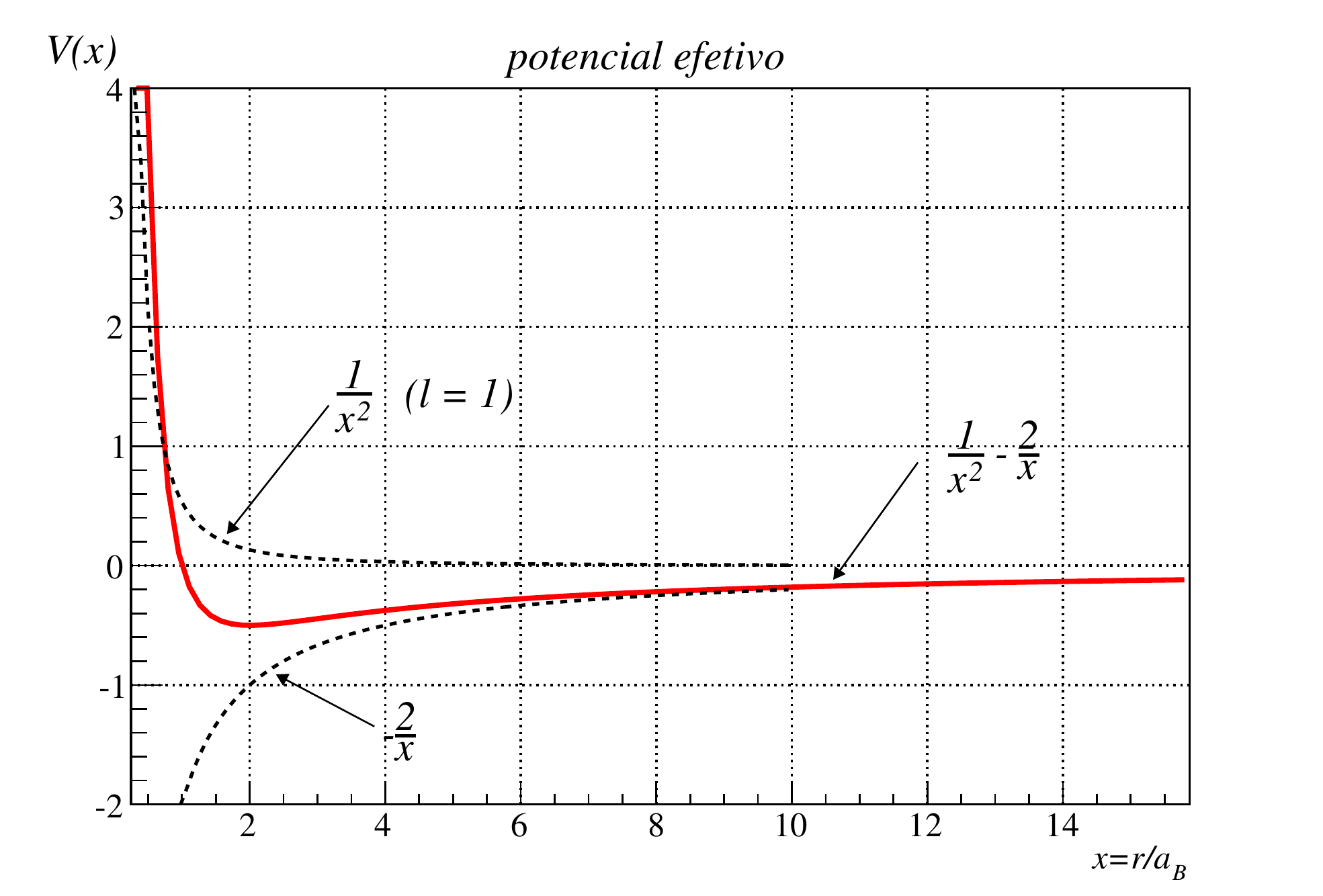}}
\vspace*{-0.3cm}
\caption{\small Potencial efetivo para o \'{a}tomo de hidrog\^{e}nio em 3 dimens\~{o}es.}
\label{pot_ef}
\end{figure}

Fazendo-se a substitui\c{c}\~{a}o de vari\'{a}vel $r = xa_{_B}$, sendo  $a_{_B}=\hbar^2/(me^2)$  o raio de Bohr, pode-se  reescrever a equa\c{c}\~{a}o diferencial
anterior, equa\c{c}\~{a}o~(\ref{eq-radial-3d}), para uma nova fun\c{c}\~{a}o $y(x) = R(r)$, como uma equa\c{c}\~{a}o de
autovalor
\begin{equation}\label{eq-radial-3d-y}
\frac{\mbox{d}^2 y}{\mbox{d} x^2} = - \frac{2}{x}
\frac{\mbox{d}y}{\mbox{d}x} -  \Big[
\epsilon  -V(x) \Big]  y(x) ,
\end{equation}
 em que $\displaystyle \epsilon = \frac{E}{e^2/(2a_{_B})}$ \ e \  $\displaystyle  V(x) = \frac{\ell(\ell+1)}{x^2} -\frac{2}{x}$ s\~{a}o, respectivamente, a energia e o chamado potencial efetivo (figura~\ref{pot_ef}), em unidades at\^{o}micas.

Nesse caso, a equa\c{c}\~{a}o que se pretende resolver pelo m\'{e}todo de Numerov apresenta um termo que envolve a derivada primeira, e pode ser expressa por
\begin{equation}\label{eq-numerov}
\psi^{\prime\prime} (x) = - p(x) \psi^\prime(x) - s(x) \psi(x) ,
\end{equation}

\noindent
onde
$$ \displaystyle
\left\{
\begin{array}{l}
\displaystyle
p(x)  = \frac{2}{x} \qquad \Rightarrow \qquad p^\prime(x) = - \frac{2}{x^2} \\
\ \\
\displaystyle s(x) = \epsilon - V(x)
\end{array}
\right.
$$

De acordo com as expans\~{o}es de Taylor, equa\c{c}\~{a}o~(\ref{taylor}), pode-se reescrever a equa\c{c}\~{a}o~(\ref{eq-numerov}) como
{\small\begin{equation}\label{eq-numerov2}
\left( 1+ \frac{\delta^2}{12} \frac{\mbox{d}^2}{\mbox{d}x^2}\right)
\psi^{\prime\prime} (x)
= - p(x) \psi^\prime(x) - s(x) \psi(x)
 - \frac{\delta^2}{12} \frac{\mbox{d}^2}{\mbox{d}x^2}
\bigg[ p(x) \psi^\prime(x) + s(x) \psi(x) \bigg] .
\end{equation}
}

De maneira similar ao caso anterior, de acordo com a equa\c{c}\~{a}o~(\ref{aprox-01}), pode-se escrever o termo do lado direito da equa\c{c}\~{a}o~(\ref{eq-numerov2}) que cont\'{e}m derivadas de ordem 2, como
{\small\begin{eqnarray}
\frac{\mbox{d}^2}{\mbox{d}x^2} \bigg[ p(x) \psi^\prime(x) + s(x) \psi(x) \bigg] &=&
 \frac{1}{\delta^2} \bigg[ p(x+\delta)
\psi^\prime(x+\delta) + s(x+\delta) \psi (x+\delta) + \nonumber \\
\nonumber \ \\
\,  & & \displaystyle \, +p(x-\delta) \psi^\prime(x-\delta) +   s(x-\delta)\psi(x-\delta) + \nonumber \\
\nonumber \ \\
\,  & & \displaystyle \,- 2 p(x) \psi^\prime(x) - 2
s(x) \psi(x) \bigg] .
\end{eqnarray}
}

Substituindo as derivadas de primeira ordem pelas aproxima\c{c}\~{o}es
\begin{equation}\label{psi-linha}
\displaystyle
\left\{
\begin{array}{l}
\displaystyle
\psi^\prime(x) = \Big[ \psi(x+\delta) - \psi(x-\delta) \Big]/(2\delta) \\
\ \\
\displaystyle
\psi^\prime(x+\delta) =  \Big[ \psi(x+\delta) - \psi(x) \Big]/\delta\\
 \ \\
\displaystyle \psi^\prime(x-\delta) =  \Big[ \psi(x) -\psi(x-\delta) \Big]/\delta\\
\end{array}
\right.
\end{equation}

\noindent obt\'{e}m-se
$$
{\small \begin{array}{l}
\displaystyle
 \frac{\mbox{d}^2}{\mbox{d}x^2} \bigg[ p(x) \psi^\prime(x) + s(x) \psi(x) \bigg]
= \frac{1}{\delta^2} \Bigg\{
\bigg[\frac{p(x+\delta) -p(x)}{\delta} + s(x+\delta)\bigg] \psi (x+\delta) + \\
 \ \\
\qquad  \displaystyle + \bigg[ \frac{p(x) - p(x-\delta)}{\delta} +
s(x-\delta)\bigg] \psi(x-\delta) + 2 \bigg[\frac{p(x-\delta) - p(x+\delta)}{2\delta}
- s(x) \bigg] \psi(x) \Bigg\} ,
\end{array}
}
$$

\noindent ou seja,
{\small\begin{eqnarray}\label{dx2-f}
\nonumber \frac{\mbox{d}^2}{\mbox{d}x^2} \bigg[ p(x) \psi^\prime(x) + s(x) \psi(x) \bigg]
 &=& \frac{1}{\delta^2} \Bigg\{
\bigg[p^\prime(x) + s(x+\delta)\bigg] \psi (x+\delta) + \bigg[p^\prime(x) +
\\
\nonumber \ \\
\,  && \displaystyle +
s(x-\delta)\bigg] \psi (x-\delta)  -2 \bigg[ p^\prime(x) + s(x)\bigg] \psi(x)
\Bigg\} .
\end{eqnarray}
}

Levando em conta que o lado esquerdo da equa\c{c}\~{a}o~(\ref{eq-numerov2}) \'{e} igual a
$$\Big[\psi(x+\delta) + \psi (x-\delta) - 2 \psi(x)\Big]/\delta^2 , $$
pode-se escrever
$$
\begin{array}{l}
\displaystyle
 \frac{\psi(x+\delta) + \psi (x-\delta) - 2 \psi(x)}{\delta^2} =
 \ - \ p(x)
\bigg[\frac{\psi(x+\delta) - \psi (x-\delta)}{2\delta} \bigg] \ - \ s(x)\psi(x) \ +  \\
\ \\
 \qquad \qquad  \displaystyle  - \frac{1}{12}
\bigg[p^\prime (x) + s(x+\delta) \bigg]\psi (x+\delta)  - \frac{1}{12}  \bigg[p^\prime (x) + s(x-\delta)
\bigg]\psi (x-\delta) +  \\
\ \\
 \qquad \qquad  \displaystyle
 + \frac{1}{6}  \bigg[p^\prime (x) + s(x) \bigg]\psi(x)
\end{array} .
$$
Reagrupando-se os termos,  e fazendo-se
\begin{equation}
\displaystyle
\left\{
\begin{array}{l}
\displaystyle
\psi (x-\delta) = \psi_0   \\
\ \\r

\displaystyle
\psi (x) =   \psi_1  \\
 \ \\
\displaystyle \psi (x+\delta) = \psi_2  \\
\end{array}
\right.
\end{equation}
\noindent
obt\'{e}m-se a equa\c{c}\~{a}o de diferen\c{c}as de Numerov para o problema, adequada \`{a} propaga\c{c}\~{a}o da solu\c{c}\~{a}o a partir do limites do intervalo de integra\c{c}\~{a}o:
\begin{equation}\label{numerov-dif}
\psi_2 = \displaystyle \frac{\displaystyle 2\Bigg\{1 - \bigg[ s (x) - \frac{p^\prime (x)}{5}
 \bigg]\frac{5\delta^2}{12}\Bigg\}\psi_1-\Bigg\{1-p (x) \frac{\delta}{2} +
\bigg[s (x-\delta)+  p^\prime(x) \bigg]\frac{\delta^2}{12} \Bigg\} \psi_0}{\displaystyle \Bigg\{1 + p(x)
\frac{\delta}{2} +  \bigg[ s (x+\delta) + p^\prime (x)  \bigg]\frac{\delta^2}{12}\Bigg\}} .
\end{equation}

A partir dessa f\'{o}rmula pode-se implementar um procedimento an\'{a}logo ao caso anterior para a constru\c{c}\~{a}o de solu\c{c}\~{o}es da equa\c{c}\~{a}o radial de Schr\"{o}dinger  no intervalo $(0, \infty)$.

\pagebreak

\begin{figure}[hbtp]
\centerline{\includegraphics[height=8.5cm]{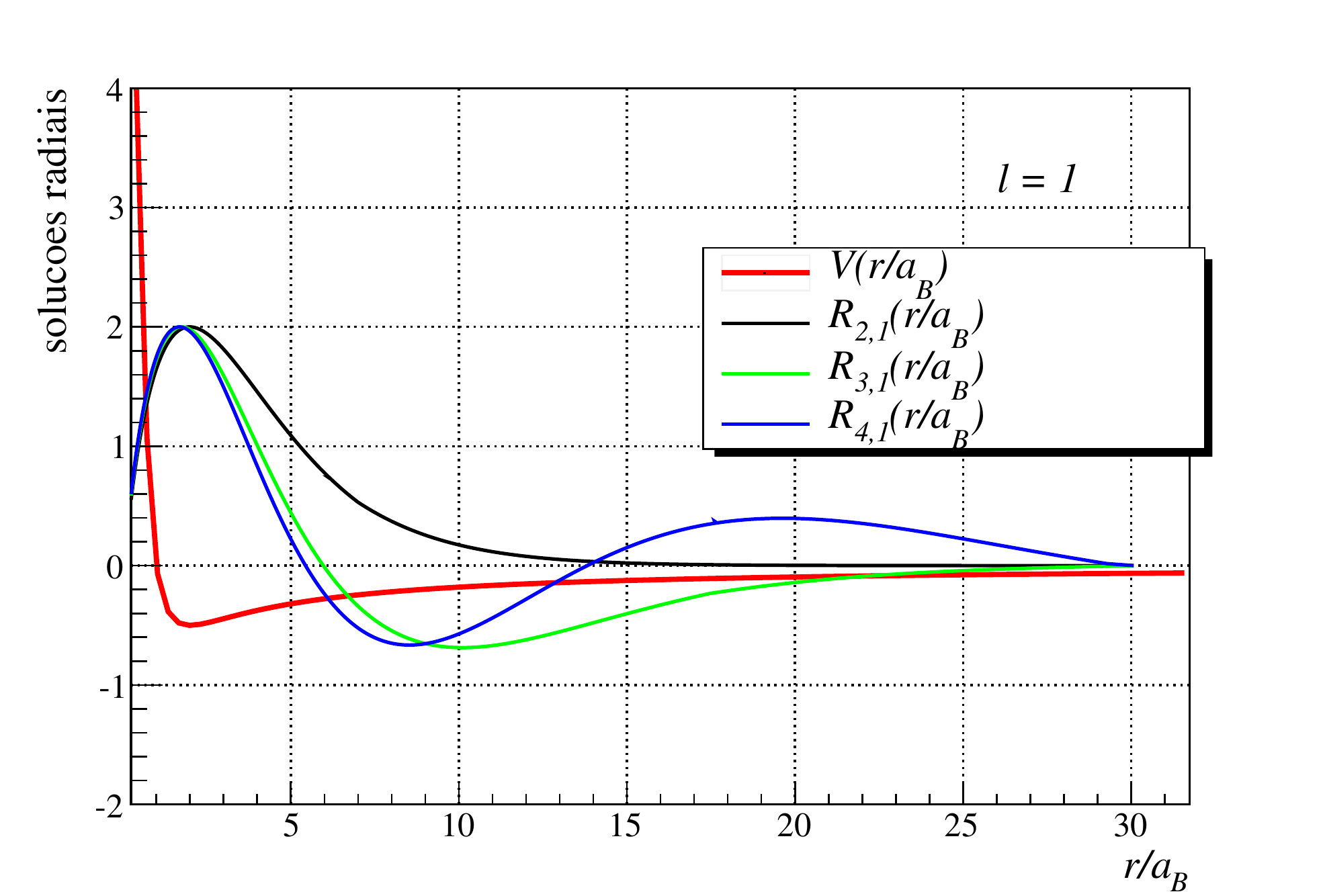}}
\vspace*{-0.3cm}
\caption{\small Potencial efetivo (em vermelho) e as solu\c{c}\~{o}es radiais  do \'{a}tomo de hidrog\^{e}nio, correspondentes aos 3 primeiros estados excitados, para $l=1$,   em 3 dimens\~{o}es.}
\label{hidro_radial}
\end{figure}

A Tabela~\ref{hidro-tab} mostra a compara\c{c}\~{a}o de alguns dos autovalores ($\epsilon$) do \'{a}tomo de hidrog\^{e}nio, calculados anal\'{\i}tica e numericamente  para $l=1$,  e a figura~\ref{hidro_radial}, as correspondentes solu\c{c}\~{o}es radiais.

\renewcommand{\arraystretch}{1.3}
\begin{table}[htbp]
  \caption{Espectro de energia dos 3 primeiros estados excitados do \'{a}tomo de hidrog\^{e}nio, para $l=1$,  em 3 dimens\~{o}es.}
  \centering
  \begin{tabular}{c|c|c}
  \hline
~~~$E(\mbox{eV})$ ~~~ & ~~~ anal\'{\i}tico~~~ & ~~~ num\'{e}rico~~~\\
\hline
  $E_2$  & -3,40 & -3,47\\
  $E_3$  & -1,51 & -1,54\\
$E_4$ & -0,85 & -0,83\\
  \hline
  \end{tabular}
  \label{hidro-tab}
\end{table}
\renewcommand{\arraystretch}{1.0}

\vspace*{-0.3cm}
\section{Coment\'{a}rio final} \label{concl}

Os programas para implementar o m\'{e}todo de Numerov foram desenvolvidos em $C^{++}$, para o compilador Cint do ROOT, vers\~{a}o~5.25/2009. Por um problema de espa\c{c}o, preferimos disponibilizar o c\'{o}digo computacional no link \url{https://dl.dropboxusercontent.com/u/8500922/numerov_program.pdf}.

Embora tenhamos apresentado apenas duas aplica\c{c}\~{o}es did\'{a}ticas, o m\'{e}todo de Numerov \'{e} suficientemente geral e robusto a ponto de poder ser usado em trabalhos cient\'{\i}ficos modernos, como a investiga\c{c}\~{a}o da depend\^{e}ncia dos n\'{\i}veis de energia do \'{a}tomo de hidrog\^{e}nio n\~{a}o-relativ\'{\i}stico com a dimensionalidade do espa\c{c}o \cite{Oguri}
ou a an\'{a}lise dos efeitos sobre a din\^{a}mica de uma part\'{\i}cula eletricamente carregada interagindo com um potencial de Chern-Simons em duas dimens\~{o}es espaciais \cite{Caruso}.

\renewcommand{\refname}{\textbf{REFER\^{E}NCIAS}} 

\end{document}